**COMMENTARY**  **Open Access**

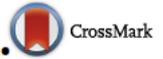

# Perspectives on software-defined networks: interviews with five leading scientists from the networking community

Daniel M Batista[6*], Gordon Blair[2], Fabio Kon[6], Raouf Boutaba[1], David Hutchison[2], Raj Jain[3], Ramachandran Ramjee[4] and Christian Esteve Rothenberg[5]

**Abstract**

Software defined Networks (SDNs) have drawn much attention both from academia and industry over the last few years. Despite the fact that underlying ideas already exist through areas such as P2P applications and active networks (e.g. virtual topologies and dynamic changes of the network via software), only now has the technology evolved to a point where it is possible to scale the implementations, which justifies the high interest in SDNs nowadays. In this article, the JISA Editors invite five leading scientists from three continents (Raouf Boutaba, David Hutchison, Raj Jain, Ramachandran Ramjee, and Christian Esteve Rothenberg) to give their opinions about what is really new in SDNs. The interviews cover whether big telecom and data center companies need to consider using SDNs, if the new paradigm is changing the way computer networks are understood and taught, and what are the open issues on the topic.

## 1 Introduction

In recent years, the network community from academia and industry has focused on the creation and management of virtual networks [1]. The motivation for virtual networks is to allow the experimentation of new protocols and new topologies, over production networks, without changing the underlying network. Through such an approach, virtual networks can be created, removed and modified, bringing to reality the possibility of easily deploying new network applications and services [2]. The procedures and mechanisms used to manage these virtual networks are part of the so-called Software-Defined Networks (SDNs) topic. Despite the fact that SDNs are already widely deployed, there are still a few number of key research questions that need to be answered [3], mainly related to some key ongoing research challenges and consideration of the impact on the previously dominant layered architectures of computer networks.

The ideas implemented in SDNs are not completely new. Approaches like overlay networks in Peer-to-Peer applications, that allow the creation of networks on the top of another network, and active networks, that allow changes on network behaviour via network packets carrying code to be executed on the routing elements, were proposed, and even implemented, a few years ago. However, SDNs have drawn more attention mainly because now the technology has evolved to a point where it is possible to scale the implementations to a level that was not possible before.

Research and Development in SDNs are well-aligned with the objectives of the Journal of Internet Services and Applications (JISA), a journal which focuses on recent advances in the state-of-the art of Internet-related Science and Technology. The Editors-in-Chief, Gordon Blair and Fabio Kon and the Assistant Editor Daniel Macêdo Batista, therefore invited the following five leading scientists, from both academy and industry, to answer five questions on the topic:

- Raouf Boutaba, Professor, David R. Cheriton School of Computer Science, University of Waterloo, Canada;
- David Hutchison, Distinguished Professor, School of Computing and Communications, Lancaster University, UK;

*Correspondence: batista@ime.usp.br
[6]University of São Paulo, São Paulo, Brazil
Full list of author information is available at the end of the article

- Raj Jain, Professor, Department of Computer Science and Engineering, Washington University in St. Louis, USA;
- Ramachandran Ramjee, Principal Researcher, Microsoft Research, India;
- Christian Esteve Rothenberg, Assistant Professor, School of Electrical and Computer Engineering, University of Campinas, Brazil, and ONF (Open Networking Foundation) Research Associate since 2013

It is an honour for us to have such major names to comment on SDN. We thank them very much in sharing with us their thoughts about this important topic.

The questions are divided into three parts:

- The first two questions focus on what is new about SDNs and what are the services and applications that can be brought to reality thanks to the research and development in the topic;
- The subsequent two questions focus on how industry is employing SDNs and how this move can affect the teaching of computer networks in Computer Science (CS) courses;
- The last question focuses on open issues in SDNs, which may be of interest for PhD students and professors.

We hope you find the next section useful and inspiring so you can advance in your academic work, improve the network performance of your organization or get ideas for new network mechanisms and services. If you have such achievements, we will be waiting for the submission of papers reporting them.

## 2 The interviews

1  *In a couple of sentences, what are Software-Defined Networks and why are they so important in Computer Science today? What have been the most significant advances in this field in recent years?*

**Raouf Boutaba:** Software-Defined Networking can be thought of as something similar to the open instruction set of x86 processor technology. It opens up the network to programmers by abstracting the implementation details of the underlying switching fabric.

The perception that SDN is particularly important in Computer Science today comes from the fact that "software" is the defining factor in SDN and software engineering is in essence a Computer Science discipline – software engineering techniques developed over the past decades can be tailored for programming the network. Having said that, traditional (before SDN) networks are in many ways run by software, but that software is embedded in hardware. Software-Defined Networking is primarily about defining the boundary between software running on proprietary hardware vs. software running on commodity servers – software that is open source, open API, and open standard, hence can be changed by network operators at will. The ability to write network control software on commodity servers is perhaps the most significant advancement in recent years.

**David Hutchison:** Software-Defined Networks can be characterised by, first, the separation of the network control plane from the forwarding process in the data plane, and second, the capability of managing a number of separate data planes from a single control plane. SDN is the latest – and arguably the most potent so far – of a series of initiatives aimed at achieving the goal of programmable networks.

Software-Defined Networks are important in Computer Science (at least in the communications and networking part of CS) because they offer the prospect of constructing networks that have improved properties, not least those of flexibility and evolvability. The most significant advances are probably in OpenFlow, a particular realization of SDN that has become popular in the research community and to some extent amongst developers; and in the creation of standards-based communities that are tackling the problematic issues of technology transfer and adoption of SDN within the industry.

**Raj Jain:** The key characteristics of SDN are programmability that allows policies to be changed on the fly. This allows orchestration (ability of manage a large number of devices), automation of operation, dynamic scaling, and service integration. Combined with virtualization, this also allows multi-tenancy and all of the above benefits to individual tenants as well.

It is important to know what SDN is not. SDN is not separation of control and data plane or centralization of control or OpenFlow. These are all one way to do SDN but alternatives are equally good. Networks can be programmed and policies can be changed without separation of data and control plane, without centralization of control, and without OpenFlow [4, 5].

**Ramachandran Ramjee:** Software-defined networks represent a fundamental shift from how networks are built and managed today. It is built upon three key changes: i) the forwarding and control functions, that are typically integrated within a router, are separated; ii) the proprietary interface between the control and forwarding functions is replaced by an open and standardized interface; and iii) the control functions that were previously distributed across all the routers in a network are logically centralized.

An analogy with the shift from mainframes to Personal Computers (PCs) is one way to think about this shift in networking. The separation of forwarding and control functions, and the replacement of the proprietary interface with a standard interface helps unbundle the router hardware from the router software which is similar to what happened with the shift from the monolithic mainframe to the disaggregated PC with different hardware and software vendors. A similar disaggregation in the router allows the router hardware to be commoditized, resulting in significantly lower cost. The control software can then be logically centralized which has recently been shown to have significant benefits in network efficiency and manageability.

**Christian Esteve Rothenberg:** We could refer to the importance of SDN in attempting to introduce missing abstractions in networking (control plane) as eloquently argued by Scott Shenker. Basically, advancing the state of the art in networking based on useful abstractions – similar to how computer science has evolved in fields like operating systems, databases and distributed systems. In terms of the networking industry, these abstractions introduce clean functional layers and APIs which allows for industry modularization, competition, and open innovation.

I would like to highlight two of my favorite aspects from SDN that make networking exciting (again). First, SDN allows more realistic (and potentially relevant/deployable) experiments to any student/researcher equipped with a single commodity laptop. From there, having standardized APIs to datapath devices (e.g. OpenFlow) and open-source controller platforms (e.g., OpenDaylight) allows running the same code in experiments using real networking environments. The vast amount of open source components in all layers of SDN [6] are to me among the most disruptive aspects of SDN, a term that is used today to encompass diverse modern networking approaches beyond the original OpenFlow split control model. Second, in operational environments, SDN abstractions through programmatic interfaces allow extensively testing the same code in a shadow environment prior to moving it to production. Allowing reproducibility of research, easing the deployability of new ideas and enhancing overall testability, in spirit of a "test-driven networking" model, these are, IMHO, gigantic steps to the field, which was highly constrained to simulation-based research experiments in academia and vendor/model-dependent not programmatic configuration interfaces in production.

2 *Could it not be claimed that SDNs emerged a long time ago through the work on active networks? In other words, is there anything actually new about this field? In convincing us, is there something that SDNs can do that active networks were incapable of? Is there any network service/application/mechanism that would be impossible without the work on SDNs?*

**Raouf Boutaba:** Though the goal of achieving network programmability is shared by active networks and SDN, their approaches are fundamentally different. The programming model of active networks (with packets carrying executable control code) is highly distributed in nature and harder to realize or secure in practice. SDN dramatically simplifies network programmability through a logically centralized control plane.

However, there is nothing conceptually new that has been introduced by SDN. SDN builds upon three key principles: a) separation of data and control planes, b) logically centralized control, and c) flow-level abstraction. All of these concepts have been around for many years.

The separation of control and data planes was extensively studied in ATM networks in the early 90's and subsequently as part of programmable networking research in mid 90's and related standardization in the late 90's (e.g., IEEE P1520 Standards). Also, in the early 90's, the Intelligent Network (IN) concept was developed by telecom operators and standardized by the ITU (ITU-T Q.1200 series recommendations) as a telephone network architecture that separates control logic from switching equipment, allowing new services to be added without having to redesign switches to support new services.

The concept of a logically centralized control plane for the network was the main contribution of policy-based networking research in the 90's where control policies are established at a logically centralized Policy-Decision Point (PDP) and communicated to network switches acting as Policy Enforcement Points (PEP) through a standard protocol (IETF COPS). Earlier, Intelligent Networks introduced the concept of Service Control Point (SCP), a logically centralized node where the service logic is located, and communicating with the switches – known as Service Switching Points (SSPs) over the standard Signaling System 7 (SS7) protocol.

Finally, SDN uses flow-level control perhaps not exactly in the same way used by prior technologies. However, the "flow abstraction" is not new and finds its similarities in existing network technologies including MPLS, COPS, RSVP, and IPv6.

**David Hutchison:** Yes, although the history is not quite as simple as that. Even before (though certainly in parallel with) the emergence of active networks, research on open network architectures and open signalling was investigating how to provide suitable and distinct control plane and data plane mechanisms. The notion and the ambition of being able to program networks to achieve particular properties arose in the early 1990s with the advent of ATM switches, multimedia-capable client devices, and the quest for Quality of Service (QoS) guarantees for end-users.

Active networks could be said to have been a flawed concept, at least as it turned out in practice... The newness of SDN is strongly connected to engineering realizations – brought about by advances in network virtualization, though that has also been around since the 1990s in association with ATM. SDN does not so much enable services/applications/mechanisms as – potentially – to endow these with the above-mentioned properties of flexibility and evolvability along with others (see 5 below).

**Raj Jain:** Active networks was the first attempt to program networks but was unsuccessful because it did not maintain the ownership boundary. A user could put a program in the packet header that could dictate how that packet would be handled and the equipment/network owner felt they would lose control and will not be able to enforce their policies. Network virtualization allows tenants to implement their own policies in a network owned by another entity and both entities can enforce their policies. The network virtualization was not available during active network days and so there was no clear separation between owners and users.

**Ramachandran Ramjee:** Yes, the fundamental idea behind SDNs, i.e., the separation of forwarding and control functions in networking has a long history. For example, the idea has been proposed in the context of integrated services digital networks (ISDN) over thirty years ago [7]. Mid-1990s also saw a spurt of research activity in the field of programmable and active networks that promised a spate of service innovations [8]. Similarly, we proposed the SoftRouter architecture ten years ago to separate the control functions from the forwarding functions of an IP router [9].

However, what distinguishes the recent surge in research interest in SDNs compared to the past is the adoption of SDN by the industry. Thus, while I don't think SDNs are by themselves a more powerful framework than prior approaches, adoption by industry galvanizes the research community to take on new and bigger challenges in the area.

**Christian Esteve Rothenberg;** Sure there are intellectual roots in SDN coming from the research work on active networks (cf. Feamster et al. [10]). Many of the ideas in SDN around programmability and virtualization could have been realized with active networks. However, I would argue that the key factor for the success of SDN is the overall maturity and technology readiness today compared to the late 90s. The enabling factors include the performance, scalability, and price factor of today's merchant silicon to implement high-speed forwarding devices combined with the processing capabilities of general purpose multi-core server technology. Who would think just a few years ago that a commodity server could forward and modify packets at 10 G speeds? These technology advancements are the reason to the success of virtual switches in overlay SDNs and are enabling the evolution towards pure software networking as being pursued by the NFV (Network Functions Virtualization) movement.

Regarding the actual new capabilities introduced by SDN and the so-called killer applications/use cases of SDN, I like to frame the discussion differently by moving the discussion from what SDN allows to how and at what cost (in the broadest sense, CAPEX, OPEX, time to market, vendor lock-in, etc.). Even if no new features were introduced by SDN but "just" a simpler network design, easier to operate, reason about, evolve, and may be most importantly to the decision makers, at a fraction of the cost, that would be more than enough arguments to be convinced about the beauties of SDN.

3. *There are some recent reports of large companies using SDN in their data centers. What are the key motivations and challenges for this and do you see a clear synergy between SDNs and data centers, particularly in large scale systems? What are the added implications of this move, for example what will this mean for traditional ISP networks?*

**Raouf Boutaba:** The key motivations for cloud owners and cloud providers to deploy SDN in their data centers are for a large part no different than those of any network infrastructure provider, i.e., reduced CAPEX and OPEX by leveraging commodity hardware and software-based controllers; improved management by means of global network view and centralized control, programmability of the data center network infrastructure through common APIs hiding networking details; flexible and rapid deployment of new data center network services, applications, control algorithms and management policies; and easier support for data center network virtualization to provide some level of network resource guarantees and flexible virtual machine migration.

Experiences from early deployments of SDN in data center networks revealed one of its main challenges that of scalability. A large-scale data center network typically involves a very large number of switching devices and an extremely large number of flows resulting in excessive flow set up and statistics gathering overheads and a performance bottleneck at the central controller. For traditional ISP networks, the scalability problem will be exacerbated as they usually deal with larger numbers of flows. Most importantly it will be challenging for ISPs using a central controller to maintain an acceptable flow set up time and a global network view given the geographically distributed locations of the network elements in a WAN. To date there has been only a few wide area SDN deployments mostly for controlling networks interconnecting geographically distributed data centers (Google's B4 is one example), but these heavily rely on aggregation of flows, which comes at the cost of a coarse control granularity. Besides the technical challenges related to the performance and scalability of SDN solutions in WANs, a major investment is required for ISPs to migrate from their currently deployed legacy hardware to SDN solutions.

Though SDN may not be deployed in traditional ISP networks in the foreseeable future, it will likely have an impact on their service infrastructure. With the current trend towards IT and telecommunications convergence, ISPs are increasingly deploying data centers at the edge of their networks to provide converged access to computing, networking and storage resources to end users. In addition, the edge cloud will be leveraged for deploying various Virtual Network Functions for content caching, security, QoS, video transcoding, etc., as well as new cloud-based access architectures and applications (e.g., C-RAN, location-aware and low latency services). In deploying these edge clouds, ISPs will likely adopt new technologies such as SDN.

**David Hutchison:** This is for companies to comment on, though it is clear that many ISPs, network operators and vendors are keenly engaged in investigating the utility of SDN aside of any data center activity.

**Raj Jain:** As indicated earlier, SDN allows orchestration, programmability, dynamic scaling, automation, visibility, performance optimization, multi-tenancy, service integration and so on. These are all the features needed to run a large data center. You cannot run a 1000+ (or even 100+) node data center without these features.

ISP networks should be able to benefit from the new technology as much as the data centers. They should use OpenDaylight to program their routers and get all the benefits. In fact, ISP industry should go the way datacenter industry has gone. That is, the ownership of equipment (cloud service provider) and the service providers (Cloud using enterprises) should be separate so that the same network can be easily used by several ISPs thus reducing the capital expenditure (CapEx) and operational expenditure (OpEx).

**Ramachandran Ramjee:** Surprisingly, the industry that is pioneering the adoption of SDNs is not the ISPs but the cloud-based services companies such as Google and Microsoft. The key reason behind SDNs appeal to these companies is the desire to reduce the cost of their massive data centers. These companies had already custom-designed servers and optimized the cooling costs. Thus, they were naturally looking for a solution to reduce their networking costs and SDNs seemed promising. The SDN disaggregation helps these companies use commodity networking switches while the centralized control of SDNs enables them to increase their average link utilization to 70 % or more (compared to 20–30 % before), resulting in significant cost savings. Once SDNs were deployed at scale in these companies, numerous other benefits of SDNs such as better network management (e.g., globally coordinate router firmware updates while being cognizant of node failures and traffic conditions) also came to light.

Traditional ISP networks were not the early adopters of SDNs because unlike the cloud-based services companies whose data center deployments were mostly greenfield, ISPs had a lot of legacy hardware and interoperability issues. Thus, the cost of migration to SDN was a major impediment and this is perhaps one of the key reasons why prior research efforts in the area did not result in commercial adoption. However, now that the benefit of SDNs have been demonstrated at scale and new interesting ideas on legacy migration to SDNs are being researched, I believe it is only a matter of time before the ISPs also migrate to the SDN architecture.

**Christian Esteve Rothenberg:** SDNs and data centers are a natural fit and non-surprisingly most commercial, in production SDNs are in the data center. Cloud-scale data centers are often green-field scenarios, that means there is not much legacy support required, which lowers the barriers to entry of SDN offerings. Data centers are single-domain silos where inter-working with the remaining Internet requires only speaking BGP. In these clean slate data center scenarios, network architects and operators have multiple choices to decide on the design where the main driver is, as usual, low cost provided it is able to scale in functionality and capacity. The winning approach seems to be a well designed physical IP network based on simple commodity devices and traditional distributed routing protocols (e.g. OSPF or BGP with ECMP) with the main goal of high-capacity IP forwarding. All virtualization features and advanced services like security or QoS are provided in the so-called overlay network implemented by programming the software-based edge virtual switches, which get all the required state and logic from a centralized control/management software (e.g., OpenStack plus SDN controller of choice such as OpenDaylight). These SDN controllers are implemented applying lessons (and open source components) from fault-tolerant, scalable distributed systems.

A natural evolution we can expect is that successful SDN models (probably starting with overlay approaches) will slowly enter ISP networks, where edge locations are the operator PoPs and Internet eXchange Points, which host and increasing amount of server racks. These computing pods will deliver not only application services such as CDN caches and Web front-ends but also networking services in virtualized software-based appliances, in contrast to current hardware-based boxes. This shift to all software implementations is being driven by operators within the NFV (Network Functions Virtualization) initiative, and SDN is likely to play an enabling role in providing programmable, dynamic connectivity between the NFV instances.

4. *SDN researchers often make bold claims about the impact of their work, particularly related to the network architecture and the ease of network administration. Is it time to re-write the textbooks to move away from layered architectures, and can we look forward to the day when we can employ network administrators that do not need a good degree in Computer Science?*

**Raouf Boutaba:** SDN is still rolling out in the data centers and has a long way to go before it can be deployed in WANs, especially in the core Internet transit network. It is too early to say whether textbooks need to be rewritten. In turn the current Internet architecture has passed the test of time. There are problems, but what about the new SDN architectures? We are incapable of being objective and need to stand back to judge their viability, which is not possible today. Similarly, we are still far away from when we will (eventually) move away from the TCP/IP stack. Besides, current SDN protocols (e.g., Openflow) operate on the TPC/IP stack implemented in all end systems and used as the basis for handling flows in the network. Changing this state of affairs will require changes in all end systems and applications at the least.

In general, layering is about separation of concerns. I am not sure to what extent implementing some functions in software (as opposed to embedding them in hardware) or implementing the service logic in a logically centralized controller instead of distributing it into the switches suggests that these functions are not needed or that separation of concerns is less relevant.

As a teacher of computer networking, I appreciate the layering in the network architecture as it helps me organize the course content, structure the lectures and have focused discussions with the students. Layering has many advantages (reduced complexity through separation of concerns and ease of maintenance through modularity). It has also disadvantages mainly in terms of performance (sometimes redundant functionality) or lack of optimized operations (particularly stresses in wireless communications where cross layer interactions/design may be useful, e.g., exploiting properties of physical channel at network or application layers). However, the advantages by far surpass the disadvantages.

With regard to network administrators, we will need to do the opposite, i.e., we will need network administrators who are also good programmers. Instead of just following a manual for a switch they will need to write program/scripts to operate and manage the network.

**David Hutchison:** No: layers should and will surely stay – they are (in)valuable pedagogical and design elements – and this will remain so in order to describe the history and developments of the communications world. Network administrators may eventually be replaced or at least assisted by software that provides the equivalent of their know-how and experience in managing networks – but such skilled people will be needed for the foreseeable future, although (perhaps, eventually) in fewer numbers.

Beware the hype! If or when SDN can fulfill its considerable promise and demonstrate its influence on new network architectures and industry practices, that will be the time to claim a real impact.

**Raj Jain:** SDNs are also layered. Instead of being non-layered, SDNs have multiple groups of layers on the top of each other. The bottom group belongs to the equipment owner. The next group belongs to the service provider and so on. The view of each group is different and so the additional APIs (application programming interfaces) have to be designed to translate the upper groups requirements in to the lower group's services.

Computers and networks are becoming easier to use. Today every person can program their smart phones but that does not mean that there is no need for administrators with a degree in computer science. In fact, the demand for computer science courses is increasing, if anything. Previously, only degree owners could use the computer but today anyone can use it. But to design and maintain these computers we need knowledge of internal workings of these computers. Similarly, SDN makes operation of datacenters easy but does not obviates the need for knowing how the equipment works and organized particularly if you want to troubleshoot problems.

**Ramachandran Ramjee:** I don't believe that we should move away from layered architectures. Layering in networking is the analog of modular programming in software design and is the key mechanism that allows independent evolution of different networking functions. However, networking as a field has seen rapid change in recent years and textbooks need to get updated. For example, the original 7-layer OSI model is better replaced by the layered model that is more widely used today. Similarly, SDN enables formal reasoning of networks that has not been emphasized earlier and this aspect may need to be introduced in newer textbooks. As regards network administrators, see my response to the next question.

**Christian Esteve Rothenberg:** Good design patterns for network architectures – be it SDN-based or traditional – are not going away. If SDN allows easing network administrations by means of automation and less error-prone configuration tasks that would mean network administrators have more time and energy to focus on higher value tasks. That's great! The better their CS degree the higher value to the business we can expect! Networking professionals that are handy with programming skills (at least at a scripting level and high-level languages like Python) are more likely to excel in their carriers. In parallel, the shift to software means opening the door to a larger community of CS professionals that may enter the networking field, provided their networking foundations are up to date. Note that by foundations I mean understanding key networking principles (lacking of a better term like science), the fundamental trade-offs (e.g., state, distribution, performance) and related mechanisms, and not necessary specifics of standardized protocols or product-specific CLI commands. Therefore, I am less convinced about the evolving/future role of vendor certification programs. As for the text books, I don't expect the need to re-write textbooks because they become useless, but certainly new books are called for to address the needs and opportunities of SDN (and NFV) in a comprehensive manner.

  5  *What are the top research challenges for SDN over the next 3–4 years? If you are advising incoming PhD students, what are the open research questions that are worth working on?*

**Raouf Boutaba:** In the near future, traffic engineering is a particularly important topic in SDN research. One important research question is how to develop algorithms that leverage SDN abstractions to make traffic engineering decisions and ultimately better utilize the network resources? Properly utilizing the network is key for reducing OPEX and increasing Network Operators adoption.

Managing the "software" in Software-Defined Networks is one of the determining factors for the success of SDN in the long run. Indeed, the SDN control plane is a software system and software is notoriously prone to bugs. New network programming models along with appropriate verification and debugging tools are needed. The "centralized" nature of the control plane introduces additional risks that need to be addressed, including failure and security vulnerabilities of the controller.

Scalability, as pointed out in the answer to question 3, is one of the main challenges uncovered from early SDN deployment experiences. Approaches leveraging multiple controllers working together in a peer-to-peer or hierarchical manner to reduce performance bottlenecks in flow processing and flow setup time will be particularly useful for large-scale wide area SDN deployments. However, in this case, maintaining a consistent global network view across all controllers is difficult and at the least costly. Strategic placement of the controllers is also relevant here.

Another timely SDN research direction would be to investigate how SDN can provide support for Network Function Virtualization (NFV) that is currently gaining significant traction in the industry. For instance, how SDN can help in steering traffic between dynamically instantiated Virtual Network Functions, and providing support for NFV service chaining?

**David Hutchison:** The biggest challenge is for SDN realizations to help prove the promise and value of the approach.

Meanwhile there are research challenges in coming up with new architectures based on SDN that lead to networks with improved properties. Notable amongst these properties are security and resilience as well as flexibility and evolvability; cost reduction in terms of CAPEX and OPEX are also significant factors in the adoption of SDN-based approaches. One interesting research topic is the realization of NFV (Network Functions Virtualization) in practice – and the role of SDN in doing so. A related challenge is how to enable very fast, scalable service provisioning. This is a specific aspect of a much broader research topic, that of autonomic network and service management, which also encompasses resilience. These themes are being investigated in new research programmes in the UK and elsewhere.

**Raj Jain:** Almost all of the traditional topics become more intense with the coming of SDN. Security is a big issue. If someone can change 10000+ computers with one command, we need extreme security on that command. Performance optimization, troubleshooting, tenant isolation, inter-cloud, wide-area network routing and optimization are some of the topics that come to mind. Applying SDN to every type of networking media – wired, wireless, optical is already a popular topic.

**Ramachandran Ramjee:** There are at least three different dimensions along which one may pursue SDN research in my view. First, one can extend the idea of applying SDN-like architectures to different types of networks, e.g., WiFi, LTE, ISP, storage, etc., and solve challenges that are unique to those domains. Second, one can address the paucity of today's network troubleshooting tools through the design of novel techniques and rich tools that help reason about networking state (analogous to tools that exist in the programming languages area). For example, can a tool automatically ensure that the invariant "no network access to a set of servers hosting sensitive data from a guest network" is never violated through a combination of appropriately setting routing tables and firewall filters? Third, in what may be considered the holy grail in networking, can we synthesize networks and networking-state automatically from a high-level set of policies as specified by the administrator? If we can address this question effectively, then network administrators of the future will perhaps need degrees in law or management rather than computer science!

**Christian Esteve Rothenberg:** Challenging research questions are present in every layer/component of an SDN stack (cf. Diego Kreutz et al. [11]), and all of them need to be addressed by the community to realize SDN at large and for the masses, not just limited to a few players like Google, Amazon, Microsoft and so on. The SDN community as a whole is working hard on filling the identified gaps. I will focus just on three research questions I am encouraging PhD students to work on. First, to work on high available, robust SDN architectures (end-to-end and considering all dimensions east-west/north-south). This includes not only deep modeling work and analysis of the theoretical limits that allow providing a fair comparison to traditional fully-distributed network architectures, but also proof of concept prototypes and experimental validation with real traffic and equipment over long periods of time. Second, working towards an OpenFlow 2.0 protocol that allows an adaptive compilation of higher-level control programs to the actual capabilities of the data plane chipsets, in whatever form factor and "instruction set" they are implemented. Third, rethink inter-domain communications assuming SDN controllers are in place. That means research on how SDN domains may "talk" to each other and allow moving beyond BGP for inter-domain routing, which could be far more expressive than current path-based dissemination of IP subnets and indirect policy hooks, including more explicit policy declaration and the exchange of additional resources (computation, storage, services) in addition to reachable IPs in a best effort manner. Software-defined eXchanges as an evolution of Internet eXchange Points are expected to become a target scenario to deploy SDN features between domains that are looking for innovative networking solutions. Along this journey hybrid SDN-IP/legacy designs will need to be developed and deployed, a step that is already happening under so-called software-defined WAN offerings. Last but not least, I encourage PhD candidates to research on SDN approaches in the context of an enabling management and control paradigm for NFV. Complementary to these advises on candidate hot topics for their research I try to help them in not getting into the trap of getting too technological and lured by buzzwords.

Take for instance one of the newest "software-defined" thinking being applied to storage (Software-Defined Storage). How much is it a term being marketized and which are the actual intellectually interesting research challenges that may be open for academic research work? This type of questions alone are a continuous challenge we all face when deciding where to look next. It is often hard to realize that some research questions are better tackled by (or at least in close collaboration with, where possible) the industry, and that some research questions are actually non-problems, since they are very likely to be solved by the industry/market on their own in the short/mid-term.

## 3 Final remarks

By the answers presented in the previous section we can conclude that the scientific and technological advances in SDNs are bringing gains for both researchers proposing new network mechanisms and protocols, since now they can make more realistic experiments with commodity hardware, and for datacenter companies, which can implement advanced traffic engineering mechanisms. In terms of teaching, we note that SDN does not make layer architectures obsolete. It is important to keep teaching and discussing this concept.

A very important observation is that network administrators with a low knowledge of programming now need to consider going back to studying, since there is a good chance that they will deal with more programming than ever to keep their network environments working efficiently.

Some key issues in the area remain open and postgraduate students can benefit from working on them: how to manage multiple SDN controllers efficiently and with security? How to effectively realize NFV with SDN? What mechanisms are needed to employ SDN to every type of networking media (wired, wireless, optical, etc…)? And, finally, a dream of all IT managers, how to use SDN to allow the fast and automatic synthesizing of networks from high-level rules?

**Competing interests**
The authors declare that they have no competing interests.

**Authors' contributions**
DB, GB and FK carried out the selection of the researchers to be invited and prepared the five questions. GB and FK made the invitations. RB, DH, RJ, RR and CR accepted the invitation and answered the questions. DB collected all the answers, contacted the researchers to address some points in their answers, and drafted the manuscript. GB and FK reviewed the entire manuscript and sent it to final comments from RB, DH, RJ, RR and CR. All authors read and approved the final manuscript.

### About the authors

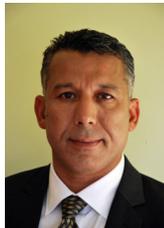

**Raouf Boutaba** received the M.Sc. and Ph.D. degrees in computer science from the University Pierre & Marie Curie, Paris, in 1990 and 1994, respectively. He is currently a professor of computer science at the University of Waterloo. His primary research interests are in network and service management. He has published extensively in these areas and received several journal and conference Best Paper Awards including the IEEE 2008 Fred W. Ellersick Prize Paper Award. He received several other recognitions such as the Premier's Research Excellence Award, the Nortel Research Excellence Award, the Nortel Excellence Award in Technology Transfer, fellowships of the Faculty of Mathematics and the David R. Cheriton School of Computer Science and several outstanding performance awards at the University of Waterloo. He has also received the IEEE Communications Society Hal Sobol Award and the IFIP Silver Core in 2007, the IEEE Communications Society Joe LociCero and the Dan Stokesbury awards in 2009, the IEEE Communications Society Salah Aidarous award in 2012, and the IEEE Canada McNaughton Gold Medal in 2014. He is the founding editor in chief of the IEEE Transactions on Network and Service Management (2007–2010), on the editorial boards of several journals, and served as general or technical program chair of a number of major IEEE conferences. He served as a distinguished lecturer for the IEEE Computer and Communications Societies. He is fellow of the IEEE, the Engineering Institute of Canada, and the Canadian Academy of Engineering.

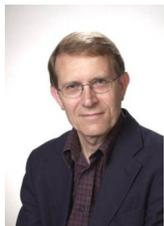

**David Hutchison** is a professor of computing at Lancaster University, and has worked in the areas of computer communications and networking for more than 25 years, recently focusing his research efforts on network resilience. He has served as member or chair of numerous TPCs (including the flagship ACM SIGCOMM and IEEE INFOCOM), and is an editor of the renowned Springer Lecture Notes in Computer Science and the Wiley Book Series in Communications, Networking and Distributed Systems.

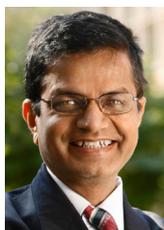

Professor **Raj Jain** is a Fellow of IEEE, a Fellow of ACM, a Fellow of AAAS, a winner of ACM SIGCOMM Test of Time award, Center for Development of Advanced Computing – Advanced Computing and Communications Society (CDAC-ACCS) Foundation Award 2009, WiMAX Forum Individual Contribution Award 2008, and ranks among the top 90 in CiteSeerX's list of Most Cited Authors in Computer Science. Dr. Jain is currently a Professor of Computer Science and Engineering at Washington University in St. Louis. Previously, he was the CTO and one of the Co-founders of Nayna Networks, Inc – a next generation telecommunications systems company in San Jose, CA. He was a Senior Consulting Engineer at Digital Equipment Corporation in Littleton, Mass and then a professor of Computer and Information Sciences at Ohio State University in Columbus, Ohio. He is the author or editor of 12 books including "Art of Computer Systems Performance Analysis", which won the 1991 "Best-Advanced How-to Book, Systems" award from Computer Press Association and "High-Performance TCP/IP: Concepts, Issues, and Solutions", published by Prentice Hall in November 2003. He is a co-editor of "Quality of Service Architectures for Wireless Networks: Performance Metrics and Management", published in April 2010. Prof. Jain has 14 patents, and has written 16 book chapters, 65+ journal and magazine papers and 110+ conference papers. His papers have been widely referenced and he is known for his research on congestion control and avoidance, traffic modeling, performance analysis, and error analysis. Google Scholar lists over 20,900+ citations to his publications. He is a co-inventor of the DECbit scheme, which has been implemented in various forms in DECnet, OSI, Frame Relay, and ATM Networks. His team has developed several switch algorithms for explicit rate-based congestion avoidance in ATM networks.

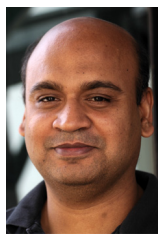

**Ramachandran Ramjee** received his B.Tech in Computer Science from IIT Madras, and his M.S./Ph.D. in Computer Science from University of Massachusetts, Amherst. He is currently a Principal Researcher at Microsoft Research, India. Previously, he spent ten years at Bell Labs as a technical manager and a distinguished member of technical staff. His research interests include network protocols and architecture, wireless networking and mobile computing. He has published over 50 papers and is a co-inventor on over 40 patents in these areas. He is the recipient of several best paper awards and the 2010 Thomas Alva Edison patent award. He has taught two graduate-level courses in wireless networks as an adjunct faculty at Columbia University. He is an ACM Distinguished Scientist and a Fellow of the IEEE.

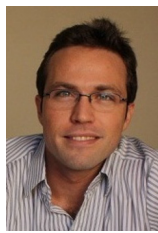

**Christian Esteve Rothenberg** is an Assistant Professor in the Faculty of Electrical and Computer Engineering at University of Campinas (UNICAMP), where he received his Ph.D. in 2010 and currently leads the INTRIG (Information & Networking Technologies Research & Innovation Group). From 2010 to 2013, he worked as Senior Research Scientist in the areas of IP systems and networking at CPqD Research and Development Center in Telecommunications (Campinas, Brazil), where he was technical lead of R&D activities in the field of OpenFlow/SDN such as the RouteFlow project, the OpenFlow 1.3 Ericsson/CPqD softswitch13, or libfluid, winner of the winning ONF Driver competition. Christian holds the Telecommunication Engineering degree from Universidad Politecnica de Madrid (ETSIT - UPM), Spain, and the M.Sc. (Dipl. Ing.) degree in Electrical Engineering and Information Technology from the Darmstadt University of Technology (TUD), Germany, 2006. Christian holds two international patents and has over 50 publications including scientific journals and top-tier networking conferences such as SIGCOMM and INFOCOM. Since April 2013, Christian is an ONF Research Associate.

### Acknowledgements
The authors would like to thank Cesar Marcondes for his comments about the questions.

### Author details
[1]University of Waterloo, Waterloo, Canada. [2]Lancaster University, Lancaster, UK. [3]Washington University in St. Louis, Saint Louis, United States. [4]Microsoft Research, Bangalore, India. [5]University of Campinas, Campinas, Brazil. [6]University of São Paulo, São Paulo, Brazil.

Received: 17 June 2015  Accepted: 27 August 2015
Published online: 30 October 2015